# TORRES FUNERARIAS *CHULLPA* EN EL VALLE DEL RÍO LAUCA: UN PRIMER ANÁLISIS ARQUEOASTRONÓMICO

# FUNERARY TOWERS *CHULLPA* IN THE LAUCA RIVER VALLEY: A FIRST ARCHAEOASTRONOMICAL ANALYSIS

**Alejandro Gangui** [1]

Universidad de Buenos Aires, Facultad de Ciencias Exactas y Naturales, Argentina.
CONICET – Universidad de Buenos Aires, Instituto de Astronomía y Física del Espacio (IAFE), Argentina

En este trabajo empleamos los métodos de la arqueoastronomía para analizar la orientación, posiblemente astronómica, de grupos numerosos de torres funerarias *chullpa*, principalmente de los siglos XII al XVI, ubicadas en el valle del río Lauca del altiplano central boliviano. A pesar de su gran relevancia histórica, tanto en las creencias y costumbres funerarias de las poblaciones locales como en la constitución del paisaje en el altiplano, poco se sabe sobre la relación de estos monumentos mortuorios con el cielo. Varios autores, desde cronistas de la época colonial hasta exploradores más modernos, indican que las torres-tumba de estas regiones se orientan de forma tal, que partes importantes de su estructura (en general, los vanos o entradas de las chullpas) apuntan hacia el levante en el horizonte oriental, de manera de impregnarse de los primeros rayos del Sol. Sin embargo, el orto solar cambia su ubicación de manera notable en diferentes épocas del año. Dada la falta de información escrita u otra forma de documentación original, para poder afirmar el uso de una orientación sistemática es preciso efectuar la medida de un número estadísticamente significativo de monumentos. Presentamos aquí los resultados del análisis de la orientación espacial precisa de los vanos de 80 torres medidas *in situ* durante un trabajo de campo en el valle del río Lauca. Hallamos que, excepto unas pocas, todas las construcciones poseen los ejes de los vanos orientados hacia oriente y dentro del rango solar, entre los acimuts extremos del movimiento anual del Sol al cruzar el horizonte local, con una concentración notable de entradas que apuntan levemente hacia el norte del este. El presente constituye el primer estudio sistemático sobre las orientaciones de las chullpas del río Lauca y puede proporcionar información crucial para comprender la evolución y el alcance del fenómeno chullpario en el altiplano boliviano y en toda la región circundante.

*Palabras clave: orientación de torres funerarias, arqueoastronomía, cultura aymara, astronomía y sociedad.*

In this work we employ the methods of archaeoastronomy to analyze the orientation, possibly astronomical, of numerous groups of *chullpa* funerary towers, mainly from the 12th to 16th centuries, located in the Lauca River valley of the central Bolivian highlands. Despite their great historical relevance, both regarding the beliefs and funerary customs of the local populations and the characteristics of the landscape in the highlands, little is known about the relationship of these mortuary monuments with the sky. Several authors, from chroniclers of the colonial era to more modern explorers, indicate that the tomb towers of these regions are oriented in such a way that important parts of their structure (in general, the entrances of the chullpas) point towards the sunrise on the eastern horizon, in order to be imbued with the first rays of the Sun. However, the sunrise changes its location noticeably at different times of the year. Given the lack of written information or other forms of original documentation, in order to affirm the use of a systematic orientation, it is necessary to measure a statistically significant number of monuments. We present here the results of the analysis of the precise spatial orientation of the entrances of 80 towers measured *in situ* during field work in the Lauca River valley. We find that, except for a few, all the buildings have the openings' axes oriented towards the east and within the solar range, between the extreme azimuths of the annual movement of the Sun as it crosses the local horizon, with a notable concentration of entrances that point slightly towards the north of due east. Our work is the first systematic study of the orientations of the chullpa towers of the Lauca River and can provide crucial information to understand the evolution and scope of the chullpa phenomenon in the Bolivian highlands and in the entire surrounding region.

*Keywords: funerary tower orientation, archaeoastronomy, aymara culture, astronomy and society.*

## I. INTRODUCCIÓN

La observación del cielo ha ocupado la atención de nuestros antepasados desde épocas muy tempranas. Al carecer de instrumentos sofisticados seguían el movimiento de los cuerpos celestes a simple vista. Reconocían los momentos singulares de los astros y la repetición cíclica de las estaciones del año. Determinaban, entre otros, las posiciones de salida y puesta del Sol en los solsticios, las de la Luna en los lunasticios, y los ortos y ocasos de las estrellas más prominentes. En reiteradas ocasiones levantaban estructuras, a veces monumentales, alineadas con esas direcciones o elegían como emplazamiento de sus lugares sagrados y tumbas, aquellos que se encontraban en un sitio singular de forma que alguno de los fenómenos descritos con anterioridad se produjese sobre una montaña sagrada o en algún otro referente topográfico importante. La relación entre paisaje celeste y paisaje terrestre, es decir, el Paisaje con mayúsculas, ha sido siempre mucho más íntima de lo que hoy en día podría parecer. La importancia de este hecho se refleja también en la reciente aprobación por parte de la UNESCO de la iniciativa "astronomía y patrimonio mundial" que trata de identificar y proteger aquellos lugares, o aquellos elementos intangibles de nuestra cultura, donde la astronomía haya jugado un papel fundamental (Ruggles y Cotte 2011).

---

[1] gangui@iafe.uba.ar

Sabemos que el paisaje, en su definición más general, incluyendo aspectos terrestres y celestes, jugó un papel relevante en la localización y la orientación de edificios en diversas culturas del pasado. Esto sucedió en latitudes muy distantes, por ejemplo, en el Egipto antiguo (Belmonte y Shaltout 2009), como así también en regiones más cercanas, siendo el imperio Inca un caso paradigmático (Bauer y Dearborn 1995). En el caso de las organizaciones sociales conocidas históricamente como Confederaciones o "Señoríos" aymara de los siglos XII al XVI, es poco lo que sabemos sobre los aspectos prácticos y cultuales de su astronomía. Estudiar la distribución y orientación de los monumentos chullpa (como expresiones arquitectónicas de costumbres mortuorias), tan relevantes en la cosmovisión andina (Nielsen 2008), nos permitirá indagar de forma novedosa sobre el posible influjo de la cultura Tiwanaku en el culto y en la construcción de tumbas y torres sepulcrales posteriores y que perduraron incluso durante las dominaciones inca y luego colonial hispánica.

En los últimos años se han multiplicado los estudios arqueoastronómicos sobre monumentos rituales y funerarios en varios sitios de relevancia arqueológica. Tal es el caso de los monumentos saharianos de piedra seca que representan importantes marcadores culturales (Gauthier 2009) y cuya distribución se ajusta muy bien con la visibilidad de objetos celestes particulares. Se ha verificado que, en el Sahara central, el patrón de orientaciones sigue con alta probabilidad la dirección de la Luna o del Sol nacientes (Gauthier 2015) y, en el caso del Sahara occidental, estudios estadísticos de varios centenares de monumentos muestran un patrón general de orientaciones hacia el horizonte oriental, principalmente agrupadas ligeramente al sur de la posición de salida más meridional de la Luna, el lunasticio mayor sur (Rodríguez-Antón et al. 2023). También en el Cuerno de África, la región oriental de ese continente donde el mar Rojo se conecta con el océano Índico, estudios recientes (Cornax Gómez et al. 2022) sugieren que una gran muestra de montículos de piedras (*cairns*), estelas y entierros antiguos en el campo de túmulos de Heis (Xiis, en Somalilandia) podrían estar orientados hacia objetivos astronómicos. Por ejemplo, estos investigadores llegan a resultados que muestran una alta concentración de orientaciones hacia el lunasticio mayor norte, donde la Luna, cuando alcanza su declinación máxima, cruza el horizonte. Dada la importancia de la Luna llena para esos pueblos nómades del desierto, la salida más extrema de las Lunas llenas de invierno a lo largo de los años, definida como la Luna llena antes y después del solsticio de invierno boreal, coincidiría con el patrón que fue encontrado en su estudio.

## II. LA CULTURA AYMARA: CONTEXTO HISTÓRICO

Luego del colapso del imperio y la cultura Tiwanaku, entre los años 1000 y 1100 d.C., aparecen en la región altiplánica boliviano-peruana varios señoríos o etnias que se disputan el territorio. Se trata de grupos aymara-parlantes ("jaqi aru", lengua de la gente o aymara) que irrumpen desde el sudoeste después de una época marcada por una extrema sequía que alteró severamente el sistema agrario manejado por el antiguo imperio (Kolata 1993). Estos reinos o señoríos se desarrollaron en el altiplano hasta aproximadamente el año 1450, cuando los incas invadieron la región.

La conquista inca de Carangas, región del Collao (o Collasuyo, así llamado por los incas) que nos ocupará en el presente trabajo, según detallan Gisbert y colaboradores (Gisbert et al. 1994; 1996), comenzó con Pachacuti y se completó durante el reinado de Tupac Inca Yupanqui. Según la crónica del Padre Bernabé Cobo en el siglo XVII, Tupac Yupanqui logró sorprender a Collas y a otras etnias (los Pacajes) en la contienda. El Inca "se encaminó al Collao detrás de las sierras de Vilcanota, y vino a salir a Chungará, tomando por las espaldas al ejército de los Collas". De acuerdo con esta cita el Inca penetró en el Collao por la zona del volcán Sajama, el río Lauca y el lago Chungará (Michel López 2021). En poco tiempo, la mayoría de los señoríos aymara fueron conquistados por los incas, ya sea por pactos de sumisión o por pérdida de batallas y la correspondiente anexión forzosa.

En el ámbito arquitectónico lo más característico de todos estos grupos aymara son sus construcciones fortificadas y sus majestuosos monumentos mortuorios. Las primeras, llamadas pucaras en quechua, fueron emplazadas en lo alto de los cerros y en lugares estratégicos para la defensa. Los segundos eran mausoleos o construcciones funerarias que, como ya señalamos, recibieron el nombre de chullpas, torres de variados tamaños, con planta cuadrada o circular y cubiertas con una bóveda por avance, en muchos casos con decoraciones muy vistosas, que se encuentran en grupos formando extensas necrópolis (Gisbert et al. 1996: 7). A las chullpas se las encuentra construidas en adobe, piedra cortada o labrada. Las más antiguas estudiadas datan, en promedio, aproximadamente del año 900 de nuestra era (Pärssinen 1993). En particular, para las torres de la región de Pacajes, este autor da una fecha radiocarbónica que delimita aproximadamente su construcción entre 1450 y 1652. Esto señala que existen en el altiplano chullpas contemporáneas a la ocupación inca y a la llegada de los españoles, como se esperaba, pues, por ejemplo, las chullpas monumentales de Sillustani (en Perú) muestran una factura similar a la característica constructiva incaica (Gisbert et al. 1996).

Hay todavía amplia discusión sobre el verdadero significado o utilidad de las chullpas en el paisaje altiplánico. Las investigaciones ya han dejado de lado la idea de que las torrecillas eran exclusivamente sepulcros de elite de personajes prominentes (*mallkus* o señores de hombres y territorios, cabezas de linaje), considerándolas más como verdaderas encarnaciones monumentales del ancestro mismo que se desea honrar. Y en tal capacidad las chullpas serían responsables de hacer lo que hacen los ancestros, es decir, "proteger los campos y los rebaños, y promover su fertilidad; proteger la cosecha; traer prosperidad a sus descendientes y proporcionarles comida, agua y otros bienes (almacenados); representar al grupo ante extraños; defender la comunidad y su territorio; luchar contra sus enemigos; inspirar decisiones políticas", y demás acciones fundamentales para la comunidad (Nielsen 2008). Además, hay evidencias de que, antes y después de la dominación inca, las chullpas eran empleadas como marcadores de límites territoriales (Fig. 1) o, de alguna manera, como construcciones que señalaban las tierras controladas por diferentes familias, linajes u organizaciones comarcales (*ayllus*) (Hyslop 1977).

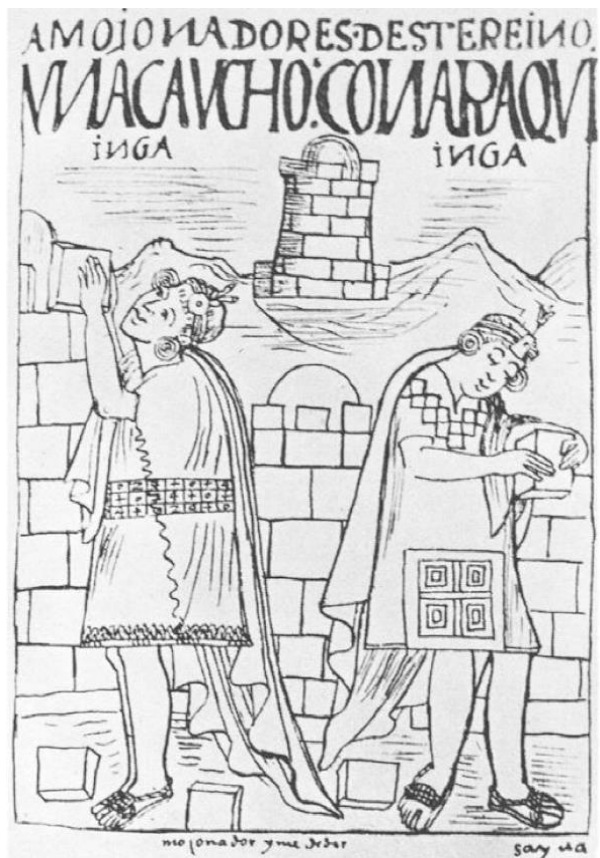

FIG. 1: *Dibujo de una chullpa como mojón de delimitación territorial inca en* Nueva crónica y buen gobierno, *de Guaman Poma de Ayala [1615].*

## III. CHULLPAS: TORRES FUNERARIAS DEL ALTIPLANO

Podemos imaginar la sorpresa de los primeros exploradores europeos al encontrarse con estas vistosas construcciones mortuorias torriformes distribuidas en la extensa llanura altiplánica. Uno de los primeros cronistas, el soldado y explorador Pedro Cieza de León, en *La crónica del Perú* (1553), relata que "tienen estos indios distintos ritos en hacer las sepulturas, porque en la provincia de Collao (como relataré en su lugar [cap. C]), las hacen en las heredades, por su orden, tan grandes como torres, unas más y otras menos, y algunas hechas de buena labor, con piedras excelentes, y tienen sus puertas que salen al nacimiento del sol, y junto a ellas (como también diré) acostumbran a hacer sus sacrificios y quemar algunas cosas, y rociar aquellos lugares con sangre de corderos o de otros animales" (Cieza, [1553, cap. LXIII] 1984: 266). Y más adelante agrega algunos detalles sobre la ubicación y las características constructivas de las chullpas: "por las vegas y llanos cerca de los pueblos estaban las sepulturas destos indios, hechas como pequeñas torres de cuatro esquinas, unas de piedra sola y otras de piedra y tierra, algunas anchas y otras angostas; en fin, como tenían la posibilidad o eran las personas que lo edificaban. Los chapiteles, algunos estaban cubiertos con paja; otros, con unas losas grandes; y parecióme que tenían las puertas estas sepulturas hacia la parte de levante" (Cieza, [1553, cap. C] 1984: 357).

Pocos años más tarde, en 1571, el cronista y encomendero Juan Polo de Ondegardo no solo describe en detalle los enterramientos de los indios, sino sus costumbres y la pervivencia de las torres chullpa (los "sepulcros de sus mayores"), las que, pese a su prohibición, seguían en uso durante la colonia castellana: "Es cosa común entre indios desenterrar secretamente los defuntos de las iglesias, o ciminterios, para enterrarlos en las Huacas, o cérros, o pampas, o en sepulturas antiguas, o en su casa, o en la del mesmo defunto, para dalles de comer y bever en sus tiempos. Y entonces beven ellos, y baylan y cantan juntando sus deudos y allegados para esto" (Polo de Ondegardo [1571] 1916: 194).

Es así que, en 1574, para evitar los continuos enterramientos en los chullpares, el virrey Toledo expide una Ordenanza que dictamina "que cada juez en su distrito haga que todas las sepulturas de torres que están en bóvedas en las montañas, e sierras, se derruequen e haga hacer un hoyo grande donde se pongan revueltos los huesos de todos los difuntos que murieron en su gentilidad" (Gisbert et al. 1994: 437). Como sabemos, estas órdenes se ejecutaron solo en parte, a juzgar por la gran cantidad de chullpares que sobrevivieron a nuestros días. Incluso varias décadas después de esta Ordenanza, las chullpas seguían atrayendo la atención de los cronistas (Fig. 2). Es el caso del ya mencionado Padre Cobo, quien, en su *Historia del Nuevo Mundo*, de 1653, escribía: "Hacíanlas por las vegas, dehesas y despoblados, unas cerca y otras lejos de sus pueblos. Todas eran en forma de torrecillas, las menores de un estado [unos 195 cm] de alto, poco más o menos, al talle de nuestras chimeneas, algo más capaces, y las mayores de cuatro a seis estados de alto. Todas tienen las puertas al oriente, y tan bajas y estrechas como bocas de horno, que no se entra en ellas sino pecho por tierra." (Cobo, [Lib. MV, cap. 18] 1964 II: 271-273).

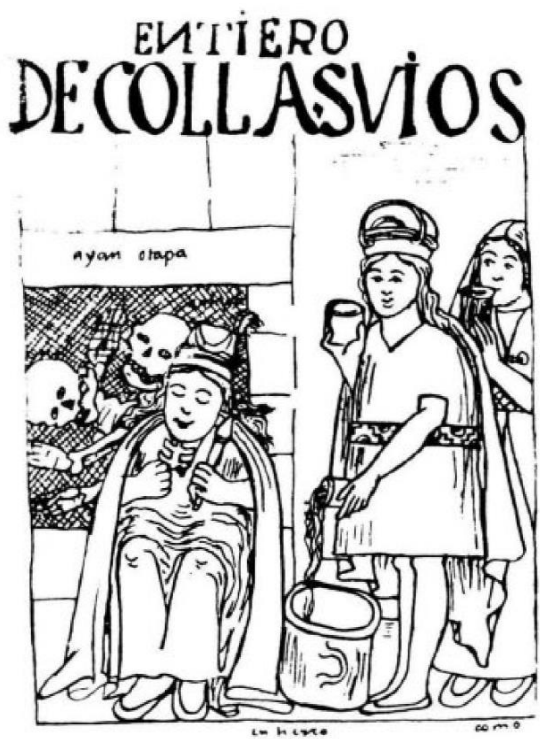

*FIG. 2: Entierro entre Collas, al fondo se ve una chullpa. Dibujo de* Nueva crónica y buen gobierno, *de Guaman Poma de Ayala [1615].*

## IV. ORIENTACIÓN DE LAS CHULLPAS: ¿ASTRONOMÍA O TOPOGRAFÍA?

Un elemento que se destaca de la narración de estos cronistas tempranos es la orientación de los vanos de entrada de las chullpas hacia el oriente: hacia el "nacimiento del sol" o "la parte de levante" según Cieza de León, o con "las puertas al oriente" según el Padre Bernabé Cobo. Esta orientación que podríamos llamar astronómica es coherente con la idea de que los cuerpos de los difuntos, guardados por la eternidad en el interior de las torrecillas, recibieran los primeros rayos del Sol de cada día, impregnándose de la energía revigorizante del astro.

Sin embargo, el paisaje andino para la cultura aymara estaba poblado de elementos sagrados y su cosmovisión incluía tres planos o "Pachas" (superior, terrestre e interior) que se conectaban a través de sitios especiales, lugares míticos de origen de donde había salido el primer ancestro, llamados pacarinas o huacas (Harris y Bouysse-Cassagne 1988: 246). Por lo tanto, es de esperar que la orientación topográfica, hacia sitios especiales del paisaje terrestre, también haya dejado su rastro en las chullpas.

Como narra Cieza de León: "cuentan estos indios que tuvieron en los tiempos pasados por cosa cierta que las ánimas que salían de los cuerpos iban a un gran lago, donde su vana creencia les hacía entender haber sido su principio" (Cieza [1553, cap. XCVII] 1984: 349). Así, lagos (como el Titicaca en el actual límite entre Bolivia y Perú) o incluso lagunas prominentes, estaban muchas veces dotados de poderes de recreación o revitalización. La misma veneración se daba con los volcanes y montes nevados que se destacan en el paisaje del altiplano andino, asociados con el culto a los antepasados. Como resultado, es natural que en el trabajo de campo de años subsiguientes se hayan encontrado chullpares orientados hacia los lagos, hacia los cerros tutelares, llamados *apus* (e.g., Reinhard 1983), o hacia los montes prominentes donde residen los antepasados, conocidos también como "abuelos" (*achachilas*) o gentiles (Harris y Bouysse-Cassagne 1988: 249).

Como lo señala Gil García (2002), las primeras observaciones planimétricas sistemáticas comenzaron a arrojar resultados en los que las torres chullpa no orientaban sus vanos hacia levante. Por ejemplo, Squier ([1877] 1974: 190-192) observó tempranamente que las chullpas de Acora (en Puno, Perú) se orientaban hacia el norte, en dirección al lago Titicaca. Años más tarde, también Ryden (1947: 343) señaló que las chullpas de la isla de Taquiri, en el Titicaca del lado boliviano, miraban todas hacia las orillas del lago y hacia el horizonte de las altas cumbres. Posteriormente, varios otros exploradores que recorrieron estas regiones del altiplano encontraron grandes necrópolis de chullpas con orientaciones topográficas diferentes, y la evidencia muchas veces mostraba que los vanos de las torres no se relacionaban con el surgir del Sol (ver Gil García 2002: 226 por más ejemplos).

Por lo que sabemos, entonces, la orientación de estos monumentos, lejos de ser aleatoria, parece seguir ciertos patrones claros: los vanos de las chullpas se orientarían hacia lagos, cursos de agua, montes o elevaciones particulares o, como comentan varios de los primeros cronistas, hacia la salida del Sol en el oriente. Como ya mencionamos, esta última es una característica de las poblaciones andinas, cuyas construcciones en general se orientaban en dirección del Sol naciente. Recordemos que para estos pueblos la orientación era muy relevante y estaba revestida de un valor simbólico (Bouysse-Cassagne 1987: 75). El este, o el oriente en general, era considerado la orientación de la vida y la fertilidad. El oeste (o el poniente), por su parte, estaba asociado con la muerte y la escasez. En particular, hasta hace no mucho tiempo, predominaban las casas de los pastores con las puertas orientadas hacia el levante, mientras que era en la dirección opuesta donde se depositaba la basura. Trabajos etnográficos de los últimos años mostraron que muchos rituales se estructuraban de acuerdo con la importancia de estas orientaciones espaciales.

Vemos entonces que la orientación de las torres chullpa fue un aspecto importante de la cultura aymara (Fig. 3). Sin embargo, el "grado de precisión" de las afirmaciones de los cronistas y exploradores posteriores no fue el principal foco de atención. Sabemos que estos pueblos no tenían forma de conocer la ubicación de los puntos cardinales más que por la posición del Sol. Pero sabemos también que el Sol, por ejemplo, en la latitud del río Lauca (donde realizaron trabajos Gisbert y colaboradores, y donde efectuamos nuestras mediciones), o en otras zonas cercanas, varía su posición en el horizonte al amanecer en un ángulo de unos 50° (entre los acimuts 65° y 115° aproximadamente) si uno lo observa entre un solsticio y el otro (entre aproximadamente el 21 de junio y el 21 de diciembre). Los arqueoastrónomos se interesan por medir, al menos de manera estadística, las orientaciones precisas de estas construcciones pues, aun a la distancia de siglos, dicen mucho sobre el culto solar de los pueblos y sobre sus actividades en diferentes momentos de su año (agrícola, religioso, etc.). Este es sin duda un tema sumamente interesante, no solo desde la temática astronómica, y es por ese motivo que hemos llevado adelante una misión de campo en una región cultural específica, Carangas, y hemos medido las orientaciones precisas de unas 80 chullpas en el ámbito del río Lauca.

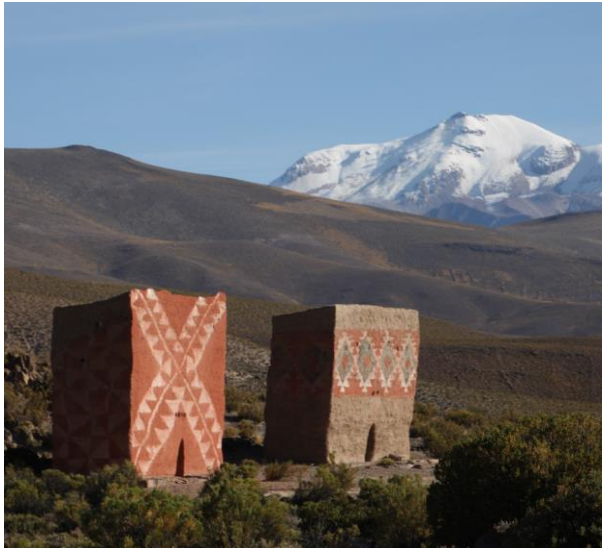

*FIG. 3: Dos torres decoradas ubicadas al sur del río Lauca. Una de ellas muestra una cruz aspada y dentada que ocupa toda la cara frontal con colores rojo y blanco, la otra decorada con un friso formado por rombos dentados de color rojo, verde, negro y blanco. La imagen fue tomada por la mañana y muestra que los vanos se abren hacia el oriente y no miran a lagunas o montes nevados prominentes. Foto del autor.*

## V. LAS CHULLPAS DEL RÍO LAUCA

La región de Carangas era uno de los señoríos de habla aymara que ocupaba la parte occidental del altiplano boliviano al oeste del lago Poopó y el río Desaguadero. Según Rivière (1982), los Carangas también dominaban gran parte de la Cordillera Occidental y territorios ubicados en el desierto de Atacama. Un corte transversal de su territorio incluyendo los enclaves de la costa y de los valles muestra para el señorío Carangas una altura media de 3800 msnm, con unos 6542 m de altura para el nevado de Sajama, volcán extinto que constituye su punto más alto, y alturas que oscilan entre 600 y 2300 m para las regiones del valle, como Lluta en la costa del Pacífico y Tiquipaya en Cochabamba (Gisbert et al. 1996: 3).

Los conjuntos de chullpas del presente estudio se concentran en medio del territorio de los Carangas, entre los sectores regidos por los nevados de Sajama y Tata Sabaya, este último al norte del salar de Coipasa, y se ubican en ambas márgenes del río Lauca (departamento de Oruro, a lo largo del piedemonte de la Cordillera Occidental) (Fig. 4). Este río nace en los lagos Chungará y Cotacotani en la provincia de Parinacota, en el actual territorio de Chile, luego ingresa en Bolivia cerca del hito fronterizo de Macaya a una altura de 3860 msnm. Esta región se caracteriza por tener un importante conjunto de chullpas, muchas de las cuales están decoradas con vistosas figuras geométricas. El Lauca penetra por una quebrada entre formaciones cordilleranas que sobrepasan los 4500 m de altura. A ambos lados del río están las lagunas Macaya y Sacabaya, al norte y al sur, respectivamente, dispuestas simétricamente respecto al Lauca "como dos grandes ojos", en un contexto antropomorfo de esa región (Gisbert et al. 1996: 22).

Estas chullpas están construidas en su mayoría con adobes planos denominados *tepes* que, en estado fresco, van conformando un entramado en los muros a la manera de la urdimbre textil. Las chullpas decoradas muestran diseños configurados geométricamente y presentan composiciones a base de figuras como el cuadrado, que girados y combinados componen formas típicamente andinas. Las más vistosas muestran listones, ajedrezados y rombos segmentados en colores rojo, blanco, negro y verde que asemejan el arte textil andino (Montero Mariscal et al. 2009: 3).

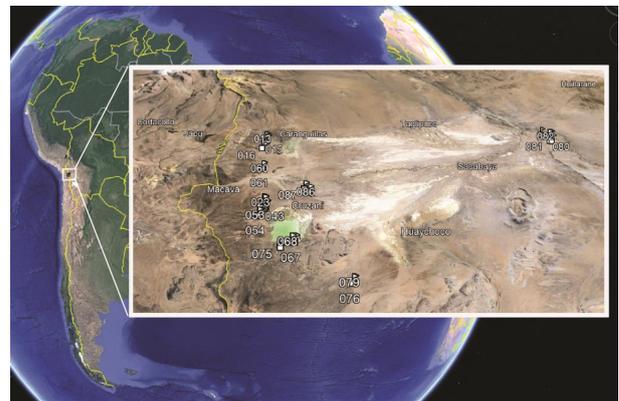

*FIG. 4: Mapa de la región de Carangas con la localización de los sitios visitados en el valle del río Lauca, entre las lagunas de Macaya (al norte) y Sacabaya (al sur). Muchas de las torres se hallan en grupos densos por lo cual no es simple individualizar cada una de ellas. El mapa incluye la ubicación precisa de las 80 chullpas georreferenciadas. Imagen sobre un mapa cortesía de Google Earth.*

## VI. MEDICIONES Y MÉTODOS DE ANÁLISIS

Obtuvimos las medidas de orientación de los vanos de las torres con brújulas de alta precisión. Los valores de la declinación magnética para distintos sitios de la región explorada oscilan entre 8°10' y 8°36' oeste (NOAA). La precisión de nuestras medidas de acimut magnético es de aproximadamente 0.5°, por lo que la diferencia en declinación magnética a lo largo del valle del Lauca entra adecuadamente dentro de nuestro error. Como una corroboración adicional, en muchos casos, especialmente con las torres más grandes y regulares, se verificaron las orientaciones medidas con imágenes fotosatelitales.

En la Fig. 5 mostramos el diagrama de orientación para las chullpas analizadas. Los valores de los acimuts consignados son los medidos para los vanos de las torres, e incluyen la corrección por declinación magnética en cada sitio particular (NOAA). Las líneas diagonales del gráfico señalan los acimuts correspondientes -en el cuadrante oriental- a los valores extremos para el Sol (acimuts de 65.3° y 115.1° -líneas continuas-, equivalente a los solsticios de invierno y verano australes, respectivamente) y para la Luna (acimuts: 59.1° y 120.8° -líneas rayadas-, equivalente a la posición de los lunasticios mayores). Recordemos brevemente que, en arqueoastronomía, con el término lunasticio mayor (norte o sur) nos referimos a las declinaciones extremas que alcanza la Luna en su movimiento a lo largo del horizonte cuando sale o se pone (ubicadas más de 5° más allá del rango de acimuts barrido por el Sol entre los solsticios).

Como vemos, casi la totalidad de las torres (75 de un total de 80 medidas) tienen los vanos orientados hacia la salida del Sol en algún momento del año, con un patrón que muestra una cierta preferencia por épocas otoñales e invernales (con acimuts hacia el norte del este geográfico), y una notoria ausencia de orientaciones cercanas al orto solar durante fechas próximas al solsticio de verano austral. Esta es una prueba de la intencionalidad astronómica que se mantuvo a lo largo de varias generaciones de comunidades aymara de esta región del señorío preincaico de Carangas.

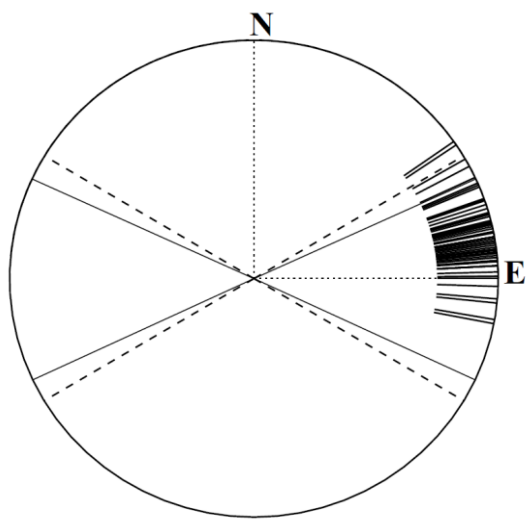

*FIG. 5: Diagrama de orientación para los vanos de las torres funerarias aymara de la región del río Lauca. Una clara tendencia a levante predomina en toda la región estudiada, con unas pocas chullpas alineadas levemente hacia el norte y fuera del rango solar.*

## VII. LA ORIENTACIÓN DE LAS CHULLPAS Y EL PAISAJE

La región cubierta en nuestro trabajo de campo es extensa (unos 300 kilómetros cuadrados, con una extensión de aproximadamente 40 km de norte a sur a lo largo del piedemonte cordillerano) y, aparte de varios montes nevados -el Sajama en la lejanía hacia el norte, los volcanes Guallatiri y Arintica del lado chileno, el Puquintica en la frontera y otros menos prominentes y más cercanos- y lagunas -por ejemplo, las ya mencionadas lagunas Macaya y Sacabaya-, no hay otros accidentes geográficos significativos que pudieran servir de referente topográfico para las construcciones analizadas. Sin embargo, como podemos ver en la Fig. 5, las ochenta chullpas medidas, sin excepción, están orientadas en un rango bastante estrecho de acimuts centrado en una dirección que apunta algo hacia el norte del punto cardinal este.

Es difícil concebir un medio por el que se hubiese podido alcanzar tal uniformidad en las orientaciones de chullpas muy lejanas a no ser que pensemos en el cielo. Sin duda nos hallamos ante un caso ejemplar de orientación astronómica en esta región de altiplano central boliviano.

Como ya señalamos, no quedan registros de los pueblos aymara que construyeron estas torres funerarias, para cuya sociedad era tan importante la memoria visual al carecer de escritura. Solo quedan las crónicas posteriores de la época colonial castellana sobre la posible orientación de los vanos. Sin embargo, el peso estadístico del patrón de orientaciones hallado parece ser suficiente para sugerir que las orientaciones se guiaban por el Sol naciente.

En tiempos del Intermedio Tardío (años 1200-1438), cuando dominaban los señoríos aymara previo al apogeo de los incas, en los diferentes ayllus de esta región del valle del Lauca, los pobladores capaces habrían estado ocupados en las labores de la tierra desde finales de la primavera y durante los meses de verano, en la producción y almacenamiento de los víveres, realizando principalmente tareas agrícolas que permitirían la futura subsistencia de la comunidad. Una vez que las cosechas hubiesen terminado, el grano ya almacenado y la tierra ya preparada para el nuevo año, la mano de obra de la comunidad habría estado disponible para embarcarse en proyectos de construcción de las necrópolis tan vistosas que, luego de más de 600 años, han llegado a nuestros días. Por ello entendemos que los vanos de las torres, si efectivamente se orientaban en la dirección de la salida del Sol en el día en que empezaba la construcción de cada monumento, deberían tener acimuts mayoritariamente equivalentes a los del orto solar en meses otoñales e invernales. Además, deberíamos hallar pocas chullpas con orientaciones cercanas al naciente del Sol en el solsticio de verano. Y esto es efectivamente lo que muestran nuestras mediciones.

Por otra parte, nuestro diagrama de la Fig. 5 no presenta acumulación de orientaciones en la vecindad de los valores extremos para la Luna en los lunasticios mayores norte o sur. Esto parecería indicar que, si bien como señala Cieza de León en *La crónica del Perú*, los aymara "tienen en cuenta el tiempo, y conocieron algunos movimientos así del Sol como de la Luna" (Cieza, [1553] 1984: 444), no hay indicios, al menos en la región de Carangas que hemos explorado, de que el orto o el ocaso de la Luna fuesen relevantes a la hora de orientar sus torres. Por último, debemos considerar la posible relevancia de aspectos climáticos propios de la región en donde se ubican las chullpas. En nuestro trabajo de campo hemos podido verificar que la mayoría de las torres medidas presentaba un fuerte grado de erosión y caída de material principalmente en su cara trasera, aquella que en general apunta hacia la cordillera y que resulta constantemente azotada por los fuertes vientos fríos y húmedos provenientes de la montaña (Fig. 6).

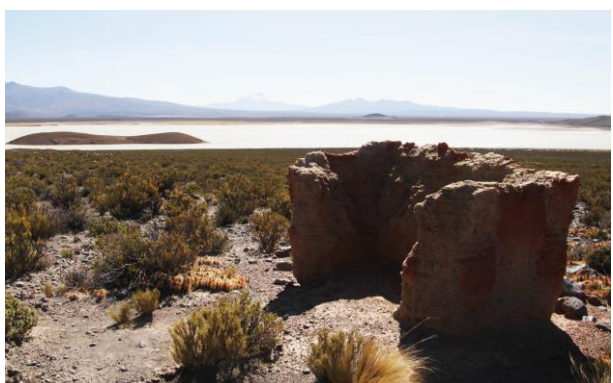

*FIG. 6: Espalda fuertemente erosionada de una chullpa decorada con cuadrados rojizos en la cercanía de la laguna e isla de Sacabaya. Como sucede a menudo, el frente y el vano permanecen en buen estado, pero la bóveda y la parte trasera están completamente derrumbadas. Foto del autor.*

Siendo esta una característica del clima propia de la región no sería extraño que los vanos de las chullpas fueran *ex profeso* ubicados a sotavento, preservando de esta manera los frentes de las torres donde en general se plasmaban los decorados más vistosos y representativos de las comunidades responsables de la construcción y manutención de los monumentos. Y no sería este el primer caso estudiado en donde el clima influye en la orientación de edificios de culto: en un contexto completamente diferente, el de las iglesias coloniales de una isla de Canarias, ya hemos comprobado que la orientación de las puertas de entrada es tal que evita enfrentarse a los vientos locales, en este caso, los alisios provenientes del norte (Gangui et al. 2016). Esta alternativa, sin embargo, requiere llevar a cabo un análisis más profundo de las condiciones climáticas en tiempos pasados, en el valle del Lauca y en diversos sitios del altiplano central con chullpares, y excede el alcance de nuestro trabajo.

## VIII. PERSPECTIVAS FUTURAS

Los pueblos aymara seguían el curso de los astros y ciertos grupos de estrellas cumplían un papel fundamental en su calendario agrícola. El mes de junio, por ejemplo, estaba marcado por varias fiestas, especialmente la de la cosecha de la papa, y el orto helíaco de las Pléyades, vistoso cúmulo estelar ubicado en la constelación de Tauro, señalaba el momento de dicha festividad. Además, según el manuscrito de Huarochirí, las Pléyades (las *Cabrillas*) tenían una función divinatoria y definían la suerte de la futura cosecha: "cuando las Cabrillas aparecen de gran tamaño, dicen: 'Este año vamos a tener maduración excelente de los frutos', pero cuando se presentan muy pequeñitas, dicen: 'Vamos a sufrir'" (Arguedas 1975, Cap. XXIX: 125; Molinié-Fioravanti 1985). En otras palabras, la tierra solo podía ser fértil si recibía la fuerza vital de las Cabrillas (*catachila* en aymara) y se lograba establecer un lazo virtuoso entre el mundo de arriba y el plano terrestre (Bouysse-Cassagne 1987: 264). Con tal presencia del movimiento de los astros enraizada en su cultura práctica y simbólica, era natural preguntarse, como lo hemos hecho aquí, si los singulares mausoleos aymara que hemos estudiado no reflejarían también una herencia astronómica en algún aspecto de su construcción.

En nuestro trabajo hemos puesto a prueba la hipótesis de la orientación solar de los vanos de un grupo numeroso de torres mortuorias ubicadas en una zona acotada del valle del río Lauca en el altiplano central de Bolivia. Hemos podido comprobar que la totalidad de las chullpas medidas se orienta con acimuts que se acomodan en el cuadrante oriental y, entre estas, prácticamente la totalidad (75 de las 80 medidas, es decir el 94% del total) orientan sus vanos dentro del rango solar (Fig. 5). Este resultado es el esperado a partir de los escritos y relatos de los primeros cronistas europeos, pero se distancia de los resultados de Pärssinen y colaboradores en la provincia Pacajes, por ejemplo, donde los investigadores hallaron que muchas chullpas estaban orientadas directamente hacia cerros y volcanes prominentes (que representan sitios sagrados en la memoria colectiva), o de los resultados de Kesseli y Pärssinen (2005) en Qiwaya, a orillas del Lago Titicaca, donde hallaron 20 chullpas, todas con orientaciones muy diferentes respecto de las del Lauca, con vanos dirigidos hacia el oeste, hacia el sur y el sudeste (Fig. 7).

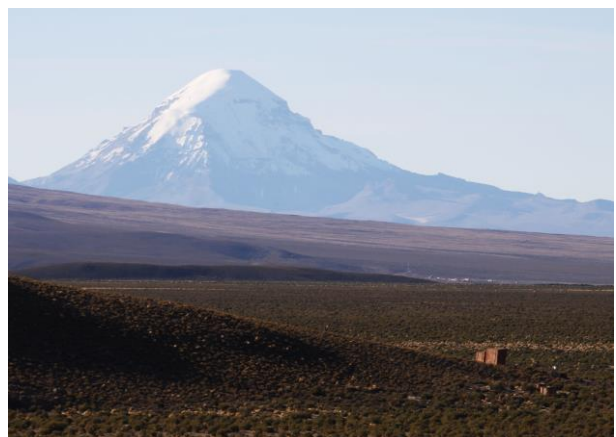

*FIG. 7: El paisaje que rodea a los chullpares incluye nevados prominentes -sitios de veneración ancestral- como el volcán Sajama de esta imagen. Sin embargo, en nuestras mediciones*

*no hay indicios de que los vanos -o incluso las espaldas- de las torres miren hacia estas montañas. Foto del autor.*

Sin embargo, el patrón específico de las orientaciones que hemos hallado aquí, donde se evidencia una alta proporción de entradas que se dirigen hacia el norte del este, con muy pocas chullpas que miran hacia el sur del este, no es por el momento simple de explicar. Como fue discutido en secciones anteriores, este resultado podría estar relacionado con aspectos de estacionalidad: simplemente, las chullpas, orientadas hacia Sol naciente, habrían sido construidas durante la parte del año en la cual las comunidades no se hallaban abocadas a los trabajos de la tierra.

Pero nuestro resultado podría deberse también a otras cuestiones, que aún debemos explorar. Hemos visto que muchos chullpares se ubican a los pies de sierras de mediana altura. Esto hace que se vean rodeados por un horizonte irregular y a veces montañoso, sobre todo en la parte trasera de las torres (en la dirección contraria hacia donde se abre el vano de las chullpas), donde el horizonte que hemos registrado en varios sitios excede los 10° de altura. Esto en general no sucede con los horizontes ubicados frente a los vanos, los cuales, como hemos visto, tienden a orientarse hacia el levante y opuestos a la cordillera. Estos últimos horizontes son bajos, y en muchas ocasiones incluso toman valores negativos debido a que las chullpas se ubican sobre pendientes y colinas (Fig. 7). En cualquier caso, sabemos que un análisis completo de nuestros datos -los que hemos reportado aquí y otros que estamos actualmente analizando- requiere tomar en cuenta la altura angular de los horizontes que rodean a las torres, pues un perfil orográfico elevado, especialmente detrás de las chullpas, sin duda cambiaría, por ejemplo, la fecha en la que el Sol en el horizonte podría alinearse con su eje.

Así, el trabajo que aún debemos realizar tiene en cuenta no solo los acimuts de los vanos sino también las alturas de los horizontes circundantes adecuadamente corregidos por refracción atmosférica. Este análisis, actualmente en progreso, nos permitirá combinar medidas locales de acimut y altura angular para obtener la declinación astronómica, coordenada ecuatorial que tiene en cuenta cómo afectan tanto la topografía local como la ubicación geográfica a la visibilidad de los objetos celestes (ver, por ejemplo, el análisis desarrollado en Muratore et al. 2023). El valor de esta coordenada estimado para una dada torre, una vez comparado con la declinación del Sol, por ejemplo (que fija aproximadamente un par de días en el año, o solo uno en el caso de los solsticios), nos permitirá verificar, entre otras cosas, si esa torre está o no orientada en una dirección que coincide con el orto solar en meses invernales, y evaluar el peso estadístico de estos resultados. Este análisis estadístico nos proveerá la distribución del número aproximado de torres por cada valor de declinación astronómica posible. Con ese dato podremos entonces verificar si la acumulación de orientaciones en el diagrama de acimuts de la Fig. 5 deja su marca también en un gráfico de declinaciones, que es, en fin de cuentas, aquello que nos señala la posible influencia astronómica -ya sea debida al movimiento del Sol o, eventualmente, a una posible orientación estelar- en la orientación de las chullpas estudiadas.

Como ya mencionamos, este análisis, al igual que una investigación más exhaustiva de los aspectos culturales relacionados con la construcción de chullpas durante la dominación del señorío Carangas, está actualmente en desarrollo y será reportado en otra ocasión.

## AGRADECIMIENTOS

## REFERENCIAS